\documentclass[aps,pra]{revtex4}
\usepackage{amsfonts}
\usepackage{amsmath}
\usepackage{amsthm}
\usepackage{amssymb}
\usepackage{graphicx}
\usepackage{subfigure}

\begin{document}

\title{Dragging spin-orbit-coupled solitons by a moving optical lattice}
\author{Hidetsugu Sakaguchi$^{1}$, Fumihide Hirano$^{1}$, and Boris A.
Malomed$^{2,3}$}
\address{$^{1}$Department of Applied Science for Electronics and Materials, Interdisciplinary Graduate School of
Engineering Sciences, Kyushu University, Kasuga, Fukuoka 816-8580, Japan}
\address{$^{2}$Department of Physical Electronics, School of Electrical Engineering,
Faculty of Engineering, and Center for Light-Matter Interaction, Tel Aviv
University, P.O. Box 39040 Tel Aviv, Israel\\
$^{3}$Instituto de Alta Investigaci\'{o}n, Universidad de Tarapac\'{a}, Casilla 7D,
Arica, Chile}

\begin{abstract}
It is known that the interplay of the spin-orbit-coupling (SOC) and
mean-field self-attraction creates stable two-dimensional (2D) solitons
(ground states) in spinor Bose-Einstein condensates. However, SOC destroys
the system's Galilean invariance, therefore moving solitons exist only in a
narrow interval of velocities, outside of which the solitons suffer
delocalization. We demonstrate that the application of a relatively weak
moving optical lattice (OL), with the 2D or quasi-1D structure, makes it
possible to greatly expand the velocity interval for stable motion of the
solitons. The stability domain in the system's parameter space is identified
by means of numerical methods. In particular, the quasi-1D OL produces a
stronger stabilizing effect than its full 2D counterpart. Some features of
the domain are explained analytically.
\end{abstract}

\maketitle


\section{Introduction}

The spin-orbit coupling (SOC) is a fundamental effect in physics of
semiconductors, induced by the interaction of the electron's spin with the
magnetic field produced by the Lorentz transform of the electrostatic field
of the crystalline lattice in the reference frame moving along with the
electron \cite{Dresselhaus,Rashba,SOC-review}. While in the solid-state
settings SOC is a complex phenomenon, it has been demonstrated that it may
be emulated in a much simpler form in binary atomic Bose-Einstein
condensates (BECs). The experiments have realized the SOC emulation in BEC
by mapping the spinor wave function of electrons into the two-component
(pseudo-spinor) wave function of the atomic condensate \cite%
{Campbell,Nature,soc2}. In terms of coupled Gross-Pitaevskii equations
(GPEs), which provide a very accurate dynamical model for BEC in the
mean-field approximation \cite{Pit}, SOC, i.e., the coupling of the momentum
and pseudospin of the matter waves, is represented by linear terms with
first spatial derivatives which mix two components of the spinor wave
function \cite{NatureRev,Review}.

Many theoretical works addressed the interplay of SOC with the intrinsic
nonlinearity of BEC, which represents, in the mean-field approximation,
effects of collisions between atoms in the dilute quantum gas. In the case
of the self-attractive sign of the nonlinearity, the analysis has predicted
the modulational instability \cite{MI} and various species of
one-dimensional (1D) solitons \cite{sol1}-\cite{Radha} under the action of
SOC. In the case of the repulsive nonlinearity, the use of spatially
periodic optical-lattice (OL) potentials has made it possible to predict 1D
gap solitons {\cite{gap-sol1,gap-sol3,He}. Otherwise, the interplay of SOC
with self-repulsion gives rise to families of dark solitons \cite{Brand}-%
\cite{dark-generation}.}

While most experimental realizations of SOC were reported in the effectively
1D geometry, SOC has also been created in the 2D setting \cite{2D-experiment}%
. Theoretical analyses of 2D setups, including SOC and the intrinsic
repulsive nonlinearity, addressed vortices \cite{vortex2}-\cite{vortex4},
monopoles \cite{monopole}, skyrmions \cite{skyrmions2,skyrmions1}, and other
delocalized states. In the presence of the OL potential and self-repulsion,
2D gap solitons were predicted too \cite{gap-sol2}. The lattice potential
can also stabilize 2D SOC solitons in the case of self-attraction \cite{we}.
Another possibility to create stable 2D solitons is offered by the
higher-order (beyond-mean-field) self-repulsion in BEC components \cite%
{droplet}, or by long-range dipole-dipole interactions \cite{DD,Raymond}.

As concerns settings based on GPEs with the mean-field cubic self-attraction
in the free 2D space, originally it was believed that all self-trapped
states generated by these models, such as Townes solitons \cite{Townes}, are
completely unstable, as the same setting gives rise to the critical collapse
which leads to destruction of solitons by perturbations \cite%
{Berge,Sulem,Fibich}. Nevertheless, it has been found that the addition of
the usual linear SOC of the Rashba \cite{Rashba} type is sufficient to
suppress the collapse and create otherwise missing ground states in the
linearly coupled system of GPEs with the cubic attractive terms \cite{Ben Li}%
. Then, the same result was produced \cite{Sherman} by the consideration of
the binary system linearly coupled by a combination of the Rashba and
Dresselhaus \cite{Dresselhaus} SOC terms. This possibility to stabilize 2D
solitons was further elaborated in Refs. \cite{Cardoso}-\cite{Kon-Sher}, see
also a brief review in \cite{EPL}.

There are two different species of 2D solitons supported by the attractive
nonlinearity in the two-component SOC system, in cases when ratio $\gamma $
of the strength of the attraction between the components to the strength of
the self-attraction in each component is $\gamma <1$ or $\gamma >1$. In the
former case, the 2D system produces stable \textit{semi-vortex} (SV)\textit{%
\ }solitons as the ground state, in which one component has zero vorticity,
and the other one carries vorticity $1$. On the other hand, the ground state
of the system with $\gamma >1$ is represented by \textit{mixed modes }(MMs),
which are composed of terms with zero and nonzero vorticities in both
components (therefore they called \textquotedblleft mixed") \cite{Ben Li}.
Simultaneously, SVs and MMs exist but are unstable at $\gamma >1$ and $%
\gamma <1$, respectively.

Mobility of solitons in SOC systems is an issue of straightforward physical
interest \cite{Beijing}. It is a nontrivial property because SOC terms break
the Galilean invariance \cite{Galil breaking,sol1,Ben Li}. By means of
numerical methods, it was found \cite{Ben Li,SKA2} that SVs and MMs may
stably move only in one direction in the 2D plane (note that SOC destroys
the system's isotropy too), with velocity $v_{0}$ taking values in a finite
interval,
\begin{equation}
0\leq v_{0}<\left( v_{\max }\right) _{\mathrm{SV,MM}}~.  \label{max}
\end{equation}%
At $v_{0}=v_{\max }$, the soliton disappears through delocalization. The
critical velocity $v_{\max }$ takes moderate values for MMs, being very
small for SVs. This observation is explained by the fact the SV's structure
is actually incompatible with the mobility, see details below.

A possibility to enhance mobility of solitons is to drag them by means of a
moving OL potential, which is an experimentally available tool \cite%
{Inguscio,Ketterle}. In this work we aim to elaborate this option and
demonstrate that the moving OL with a relatively small amplitude is able to
stabilize the motion of the 2D solitons up to much higher values of the
velocity than in the free space. The model and a relevant analytical
framework are presented in Section II. Systematic results, obtained,
chiefly, by means of numerical methods are reported in Section III. In
particular, for the dragged SVs, the increase of $v_{0}$ leads, first, to
their transformation into MMs, while the delocalization takes place at much
higher velocities. Considering both the full 2D lattice and its quasi-1D
counterpart, we conclude that the latter one produces an \emph{essentially
stronger} stabilizing effect on the moving solitons than the full 2D
lattice. Moreover, a surprising result is that the strongest stabilization
is provided by the quasi-1D lattice with the wave vector directed \emph{%
perpendicular to the velocity}. This finding is explained by the fact that
such a lattice suppresses delocalization of the moving soliton in the
transverse direction. The paper is concluded by Section IV.

\section{The model and analytical framework}

\subsection{The basic equations}

Following Ref. \cite{Ben Li}, we consider the system of coupled GPEs for two
components $\phi _{\pm }\left( x,y,t\right) $ of the BEC\ spinor wave
function written in the laboratory reference frame. The equations include,
as the first option, a square-shaped OL potential $U_{\mathrm{2D}}(x,y;t)$
with amplitude $U_{0}$ and wavenumbers $q$, moving at velocity $v_{0}$ along
the $y$ direction in the $(x,y)$ plane:
\begin{gather}
i\frac{\partial \phi _{+}}{\partial t}=-\frac{1}{2}\nabla ^{2}\phi
_{+}-\left( \left\vert \phi _{+}\right\vert ^{2}+\gamma \left\vert \phi
_{-}\right\vert ^{2}\right) \phi _{+}  \notag \\
-U_{\mathrm{2D}}(x,y;t)\phi _{+}+\lambda \left( \frac{\partial \phi _{-}}{%
\partial x}-i\frac{\partial \phi _{-}}{\partial y}\right) ,  \label{phi+} \\
i\frac{\partial \phi _{-}}{\partial t}=-\frac{1}{2}\nabla ^{2}\phi
_{-}-\left( \left\vert \phi _{-}\right\vert ^{2}+\gamma \left\vert \phi
_{+}\right\vert ^{2}\right) \phi _{-}  \notag \\
-U_{\mathrm{2D}}(x,y;t)\phi _{-}-\lambda \left( \frac{\partial \phi _{+}}{%
\partial x}+i\frac{\partial \phi _{+}}{\partial y}\right) ,  \label{phi-} \\
U_{\mathrm{2D}}(x,y;t)=U_{0}\cos \left( qx\right) \cos \left(
q(y-v_{0}t)\right) .  \label{2D}
\end{gather}%
In this notation, $\hbar $ and the atomic mass, as well as the effective
coefficient of the self-attraction in each component, are scaled to be $1$,
while $\gamma \geq 0$ is the above-mentioned ratio of the cross/self
interaction strengths, and $\lambda >0$ is the real coefficient of SOC of
the Rashba type.

Parallel to the full 2D potential (\ref{2D}), we consider its quasi-1D
variants, with the wave vector oriented parallel or perpendicular to the
velocity:%
\begin{equation}
U_{\mathrm{1D}}(y;t)=U_{0}\cos \left( q(y-v_{0}t)\right) ;U_{\mathrm{1D}%
}(x)=U_{0}\cos \left( qx\right) .  \label{Q1D}
\end{equation}%
In the latter case, the quasi-1D OL is not actually moving (but the solitons
will move in the $y$ direction), therefore Eqs. (\ref{phi+}) and (\ref{phi-}%
) with potential $U_{\mathrm{1D}}(x)$ do not explicitly depend on time.
Using the remaining scaling invariance of Eqs. (\ref{phi+}) and (\ref{phi-})
(which includes a freedom of rescaling the coordinated by an arbitrary
factor), in most cases we fix
\begin{equation}
q=2\pi /3  \label{q}
\end{equation}%
in potentials (\ref{2D}) and (\ref{Q1D}), which is a value convenient for
numerical simulations. Nevertheless, some results for other values of $q$
are presented below too, see Eqs. (\ref{5/6}) and (\ref{1/2}). Including
results in this form is relevant because in the experiment it is possible to
change the period of the OL, keeping other parameters fixed \cite{Oberthaler}%
.

It is relevant to mention that, in the framework of the usual GPE with cubic
self-attraction, 2D solitons may be stabilized not only by the full 2D
spatially periodic potential \cite{BBB,YM,YM2,Thaw}, but also by its
quasi-1D version \cite{BBB-quasi1D}, see also Ref. \cite{Barcelona-quasi1D}.
Furthermore, in the free 2D space, solitons can be stabilized by the SOC
terms taken in an essentially quasi-1D form \cite{Kon-Sher}. Somewhat
surprisingly, the present analysis reveals, in the next section, that the
quasi-1D potentials (\ref{Q1D}), especially $U_{\mathrm{1D}}(x$), provide an
essentially stronger stabilizing effect for moving solitons than the full 2D
potential.

In the moving reference frame with coordinate $\tilde{y}=y-v_{0}t$, Eqs.~(%
\ref{phi+}) and (\ref{phi-}) are rewritten as
\begin{gather}
i\frac{\partial \phi _{+}}{\partial t}-iv_{0}\frac{\partial \phi _{+}}{%
\partial \tilde{y}}=-\frac{1}{2}\nabla ^{2}\phi _{+}-\left( \left\vert \phi
_{+}\right\vert ^{2}+\gamma \left\vert \phi _{-}\right\vert ^{2}\right) \phi
_{+}  \notag \\
-U_{\mathrm{2D}}(x,\tilde{y};t)\phi _{+}+\lambda \left( \frac{\partial \phi
_{-}}{\partial x}-i\frac{\partial \phi _{-}}{\partial \tilde{y}}\right) ,
\label{phi+2} \\
i\frac{\partial \phi _{-}}{\partial t}-iv_{0}\frac{\partial \phi _{-}}{%
\partial \tilde{y}}=-\frac{1}{2}\nabla ^{2}\phi _{-}-\left( \left\vert \phi
_{-}\right\vert ^{2}+\gamma \left\vert \phi _{+}\right\vert ^{2}\right) \phi
_{-}  \notag \\
-U_{\mathrm{2D}}(x,\tilde{y};t)\phi _{-}-\lambda \left( \frac{\partial \phi
_{+}}{\partial x}+i\frac{\partial \phi _{+}}{\partial \tilde{y}}\right) ,
\label{phi-2}
\end{gather}%
\begin{gather}
U_{\mathrm{2D}}(x,\tilde{y})=U_{0}\cos \left( qx\right) \cos \left( q\tilde{y%
}\right) ,  \label{tilde} \\
U_{\mathrm{1D}}(\tilde{y})=U_{0}\cos \left( q\tilde{y}\right) ;U_{\mathrm{1D}%
}(x)=U_{0}\cos \left( qx\right) .  \label{tilde1D}
\end{gather}

In particular, Eqs. (\ref{phi+2}) and (\ref{phi-2}) with $U_{\mathrm{2D}}=0$
admit a family of continuous-wave (CW) solutions with arbitrary amplitude $A$
and wavenumber $k_{y}$ (for the definiteness' sake, we here assume $k_{y}>0$%
):%
\begin{gather}
\left( \phi _{\pm }\left( \tilde{y},t\right) \right) _{\mathrm{CW}}=\pm
A\exp \left( ik_{y}\tilde{y}-i\mu _{\mathrm{CW}}t\right) ,  \notag \\
\mu _{\mathrm{CW}}=k_{y}^{2}/2-k_{y}\left( \lambda +v_{0}\right) -A^{2}.
\label{CW}
\end{gather}%
Signs $\pm $ in front of the components of this solution are chosen\ to
select the CW branch with lower energy.

Soliton solutions to Eqs. (\ref{phi+2}) and (\ref{phi-2}) with the OL
potential (\ref{tilde}) or (\ref{tilde1D}) and chemical potential $\mu $ are
looked for in the usual form,%
\begin{equation}
\phi _{\pm }\left( x,y;t\right) =u\left( x,y\right) \exp \left( -i\mu
t\right) .  \label{umu}
\end{equation}%
These solutions, along with the respective values of $\mu $, were obtained
by means of the imaginary-time evolution method \cite%
{Tosi,imaginary,imaginary2} applied to Eqs. (\ref{phi+2}) and (\ref{phi-2})
for a fixed value of the total norm,%
\begin{equation}
N=\int \int \left( \left\vert \phi _{+}(x,y)\right\vert ^{2}+\left\vert \phi
_{-}(x,y)\right\vert ^{2}\right) dxdy,  \label{norm}
\end{equation}%
as the method is adjusted for finding solutions under this condition.

Even if the above equations are not invariant with respect to the Galilean
transform, for analytical considerations it is useful to rewrite them in the
moving reference frame, applying the formal Galilean boost to Eqs. (\ref%
{phi+2}) and (\ref{phi-2}):%
\begin{equation}
\phi _{\pm }\equiv \exp \left( iv_{0}\tilde{y}+\frac{i}{2}v_{0}^{2}t\right)
\tilde{\phi}\left( x,\tilde{y},t\right) .  \label{boost}
\end{equation}%
The accordingly transformed system is
\begin{gather}
i\frac{\partial \tilde{\phi}_{+}}{\partial t}=-\frac{1}{2}\nabla ^{2}\tilde{%
\phi}_{+}-\left( \left\vert \tilde{\phi}_{+}\right\vert ^{2}+\gamma
\left\vert \tilde{\phi}_{-}\right\vert ^{2}\right) \tilde{\phi}_{+}  \notag
\\
-U_{\mathrm{2D}}(x,\tilde{y};t)\tilde{\phi}_{+}+\lambda \left( \frac{%
\partial \tilde{\phi}_{-}}{\partial x}-i\frac{\partial \tilde{\phi}_{-}}{%
\partial \tilde{y}}\right) +\lambda v_{0}\tilde{\phi}_{-},  \label{tilde+} \\
i\frac{\partial \tilde{\phi}_{-}}{\partial t}=-\frac{1}{2}\nabla ^{2}\tilde{%
\phi}_{-}-\left( \left\vert \tilde{\phi}_{-}\right\vert ^{2}+\gamma
\left\vert \tilde{\phi}_{+}\right\vert ^{2}\right) \tilde{\phi}_{-}  \notag
\\
-U_{\mathrm{2D}}(x,\tilde{y};t)\tilde{\phi}_{-}-\lambda \left( \frac{%
\partial \tilde{\phi}_{+}}{\partial x}+i\frac{\partial \tilde{\phi}_{+}}{%
\partial \tilde{y}}\right) +\lambda v_{0}\tilde{\phi}_{+},  \label{tilde-}
\end{gather}%
Unlike Eqs. (\ref{phi+2}) and (\ref{phi-2}), this system includes direct
inter-component mixing with coefficient $\lambda v_{0}$, but does not
include the group-velocity terms, $-iv_{0}\partial \phi _{\pm }/\partial
\tilde{y}$.

\subsection{Analytical estimates}

Knowledge of the spectrum of the linearized set of equations (\ref{tilde+})
and (\ref{tilde-}) without the potential ($U_{0}=0$) helps one to predict
the existence region for solitons. A straightforward calculation for
small-amplitude excitations, taken as
\begin{equation}
\tilde{\phi}_{\pm }\sim \exp \left( ik_{x}x+ik_{y}y-i\tilde{\mu}t\right) ,
\label{excit}
\end{equation}%
yields two branches of the dispersion relation between chemical potential $%
\mu $ and wave vector $\left( k_{x},k_{y}\right) $,%
\begin{equation}
\tilde{\mu}=\frac{1}{2}k^{2}\pm \lambda \sqrt{k_{x}^{2}+\left(
k_{y}+v_{0}\right) ^{2}.}  \label{mu}
\end{equation}%
Expression (\ref{mu}) takes values in the propagation band,
\begin{equation}
\tilde{\mu}\geq \tilde{\mu}_{\min }\equiv -\lambda ^{2}/2-|\lambda v_{0}|,
\label{band}
\end{equation}%
while solitons may populate the remaining semi-infinite bandgap, $\tilde{\mu}%
<\tilde{\mu}_{\min }$. Note that the increase of the velocity pushes $\mu
_{\min }$ down, thus reducing the bandgap. All the soliton solutions
produced indeed satisfy the condition $\tilde{\mu}<\tilde{\mu}_{\min }$.

It is relevant to mention that the 1D limit of Eqs. (\ref{tilde+}), (\ref%
{tilde-}) and (\ref{tilde1D}), corresponding to no $x$ dependence and
potential $U_{0}\cos \left( q\tilde{y}\right) $, admits an obvious
substitution,%
\begin{equation}
\tilde{\phi}_{\pm }\left( \tilde{y},t\right) =\exp \left( -i\lambda \tilde{y}%
+i\left( \lambda ^{2}/2-\lambda v_{0}\right) t\right) \phi (\tilde{y},t),
\label{phi}
\end{equation}%
with which the system reduces to the singe equation:%
\begin{equation}
i\frac{\partial \phi }{\partial t}=-\frac{1}{2}\frac{\partial ^{2}\phi }{%
\partial \tilde{y}^{2}}-\left( 1+\gamma \right) |\phi |^{2}\phi -U_{0}\cos
\left( q\tilde{y}\right) \phi .  \label{single}
\end{equation}%
Evidently, Eq. (\ref{single}) is the usual 1D GPE with the standard OL
potential, which always has soliton solutions \cite{Oberthaler}, hence this
1D limit, unlike the full 2D system, admits the motion of solitons with an
unlimited velocity.

Coming back to the 2D system, a crude explanation for the existence of the
limit value of the velocity, $v_{\max }$ (see Eq. (\ref{max})), may be
proposed, based on the numerical observation that, close to the
delocalization transition at $v_{0}=v_{\max }$, the solitons are, quite
naturally, very broad in the $x$ direction, keeping weakly separated maxima
in the two components (see Fig. \ref{fig1} below), i.e., they feature
splitting of the components. This observation suggests to address a
possibility of the splitting in terms of the stationary version of Eqs. (\ref%
{tilde+}) and (\ref{tilde-}) for constant-amplitude solutions with chemical
potential $\tilde{\mu}<0$:
\begin{equation}
\tilde{\phi}_{\pm }=\tilde{u}_{\pm }\exp \left( -i\tilde{\mu}t\right) ,
\label{u}
\end{equation}%
where amplitudes $\tilde{u}_{\pm }$ may be real, and (close to the center) $%
U_{\mathrm{2D}}\left( x,\tilde{y}\right) $ is replaced by $U_{0}$:
\begin{gather}
\left( \tilde{\mu}+U_{0}\right) \tilde{u}_{+}+\left( \tilde{u}%
_{+}^{2}+\gamma \tilde{u}_{-}^{2}\right) \tilde{u}_{+}-\lambda v_{0}\tilde{u}%
_{-}=0,  \label{u+} \\
\left( \tilde{\mu}+U_{0}\right) \tilde{u}_{-}+\left( \tilde{u}%
_{-}^{2}+\gamma \tilde{u}_{+}^{2}\right) \tilde{u}_{-}-\lambda v_{0}\tilde{u}%
_{+}=0.  \label{u-}
\end{gather}%
Then, it is relevant to look for a critical point at which a solution with
an infinitesimal splitting between the components, $\Delta \tilde{u}\equiv
\tilde{u}_{+}-\tilde{u}_{-}$, appears on top of the obvious solution to Eqs.
(\ref{u+}) and (\ref{u-}) with identical (unsplit) components,
\begin{equation}
\tilde{u}_{\pm }^{2}=-\left( \tilde{\mu}+U_{0}-\lambda v_{0}\right) /\left(
1+\gamma \right) .  \label{u^2}
\end{equation}%
A simple calculation, based on equations (\ref{u+}) and (\ref{u-})
linearized with respect to $\Delta \tilde{u}$, yields the value of $v_{0}$
at the critical point:%
\begin{equation}
v_{\mathrm{\max }}=\left( 1-\gamma \right) \left( \tilde{\mu}+U_{0}\right)
/\left( 2\lambda \right) ,  \label{approx}
\end{equation}%
the splitting being impossible at $v_{0}>v_{\max }$. Finally, the
substitution of value (\ref{approx}) in Eq. (\ref{u^2}) yields the
background amplitude at the critical point, $\tilde{u}_{\pm }^{2}=-\left(
\tilde{\mu}+U_{0}\right) /2$. Because the delocalization transition proceeds
via small-amplitude solitons (see Fig. \ref{fig1} below), the present
consideration is relevant for small $\left\vert \tilde{\mu}+U_{0}\right\vert
$ and, naturally, for small values of the SOC strength, $\lambda $, to
justify the use of the constant-amplitude solution. The prediction of the
delocalization point, given by Eq. (\ref{approx}), makes sense at $\gamma <1$%
, and it does not apply at $\gamma >1$, when the formal expression predicts
a negative velocity.

This consideration reveals the possibility of the splitting between the
components which is only qualitatively similar to what occurs in the
solitons of the MM type. Therefore, it is relevant to compare dependence $v_{%
\mathrm{\max }}=\mathrm{const}\cdot \lambda ^{-1}$, predicted by Eq. (\ref%
{approx}), with numerical results (see the dashed hyperbola in Fig. \ref%
{fig3}(b) below), fitting $\mathrm{const}$ to the numerical data, rather
than using the coefficient from Eq. (\ref{approx}).

In the limit of large $\lambda $, opposite to one considered above, $v_{\max
}$ can be estimated using the variational approximation (VA). To simplify
the matters, one may apply it to the 1D limit of Eqs. (\ref{tilde+}) and (%
\ref{tilde-}), in which the $y$ derivatives are dropped (this limit case is
opposite to one considered above in the form of Eqs. (\ref{phi}) and (\ref%
{single}), where the $x$ derivatives were omitted). The stationary version
of these equations for real functions $\tilde{u}_{\pm }(x)$, defined as in
Eq. (\ref{u}), is%
\begin{gather}
\left[ \tilde{\mu}+U_{\mathrm{1D}}(x)+\frac{1}{2}\frac{d^{2}}{dx^{2}}+\tilde{%
u}_{+}^{2}+\gamma \tilde{u}_{-}^{2}\right] \tilde{u}_{+}+\lambda \left(
v_{0}-\frac{d}{dx}\right) \tilde{u}_{-}=0,  \label{u+-} \\
\left[ \tilde{\mu}+U_{\mathrm{1D}}(x)+\frac{1}{2}\frac{d^{2}}{dx^{2}}+\tilde{%
u}_{-}^{2}+\gamma \tilde{u}_{+}^{2}\right] \tilde{u}_{-}+\lambda \left(
v_{0}+\frac{d}{dx}\right) \tilde{u}_{+}=0,  \label{u-+}
\end{gather}%
Because, in the case of large $\lambda $, the delocalization always happens
with states of the MM type, one may use the following Gaussian ansatz for
mirror-symmetric components of the wave function:%
\begin{equation}
\left( \tilde{u}_{\pm }(x)\right) _{\mathrm{ans}}=A\exp \left( -\frac{\left(
x\mp \xi \right) ^{2}}{2W^{2}}\right) ,  \label{xi}
\end{equation}%
where $A$ and $W$ are the amplitude and width, $2\xi $ being the separation
between the split peaks of the components. The full form of VA turns out to
be cumbersome, but, in the limit of large $\lambda $, the prediction for $%
\left( v_{0}\right) _{\max }$, as the critical value at which the solution
for $W$, with a fixed value of the norm, becomes impossible, is simple: $%
\left( v_{0}\right) _{\max }=\lambda $ (this approximation neglects the
presence of the potential). The particular coefficient in this expression
depends on the assumptions adopted to apply VA, but the linear form of the
dependence,%
\begin{equation}
\left( v_{0}\right) _{\max }=\mathrm{const}\cdot \lambda ,  \label{linear}
\end{equation}%
is a corollary of scaling properties of Eqs. (\ref{tilde+}) and (\ref{tilde-}%
), in the absence of the external potential (the scaling leaves the total
norm (\ref{norm}) of the 2D system invariant). The scaling is corroborated
in detail below by a typical numerical solution displayed in Fig. \ref{fig7}%
. Equation (\ref{linear}) explains, approximately or exactly, numerical
findings presented below in Figs. \ref{fig3}(a) and \ref{fig6}.

We note, in passing, that the linearized version of Eqs. (\ref{u+-}) and (%
\ref{u-+}) with the 1D potential (\ref{tilde1D}) admits a parametric
resonance, accounted for by solutions in the form of
\begin{equation}
\tilde{u}_{\pm }(x)=a_{\pm }\cos \left( \frac{q}{2}x\right) +b_{\pm }\sin
\left( \frac{q}{2}x)\right)  \label{ab}
\end{equation}%
Straightforward consideration of Eqs. (\ref{u+-}) and (\ref{u-+})
demonstrates that the parametric resonance takes place at%
\begin{equation}
\tilde{\mu}=\frac{q^{2}}{2}\pm \sqrt{\left( \lambda \frac{q}{2}\right)
^{2}+\left( \lambda v_{0}\pm \frac{U_{0}}{2}\right) ^{2}},  \label{res}
\end{equation}%
where the two signs $\pm $ are mutually independent. However, the parametric
resonance does not play an essential role in this work.

\section{Results}

As mentioned above, stationary 2D soliton solutions were produced by means
of the imaginary-time integration of Eqs.~(\ref{phi+2}) and (\ref{phi-2}),
performed for a fixed total norm (\ref{norm}) of the solitons. It was then
verified by systematic simulations of perturbed evolution of the solitons in
real time that they are stable in the respective existence intervals (\ref%
{max}). The identification of $v_{\max }$ is the main objective of the
numerical analysis. For the system of Eqs.~(\ref{phi+2}) and (\ref{phi-2})
with $\gamma =0$ and $\gamma =2$, we report the results with $N=5$ and $3$,
respectively, as these values make it possible to present generic results.
Numerical computations were performed in the 2D domain of size $12\times 12$%
, which is sufficient to display all details of the 2D soliton profiles.

\subsection{Dragging 2D solitons by the square-shaped OL in the absence of
the nonlinear cross-interaction ($\protect\gamma =0$)}

\begin{figure}[h]
\begin{center}
\includegraphics[height=6.cm]{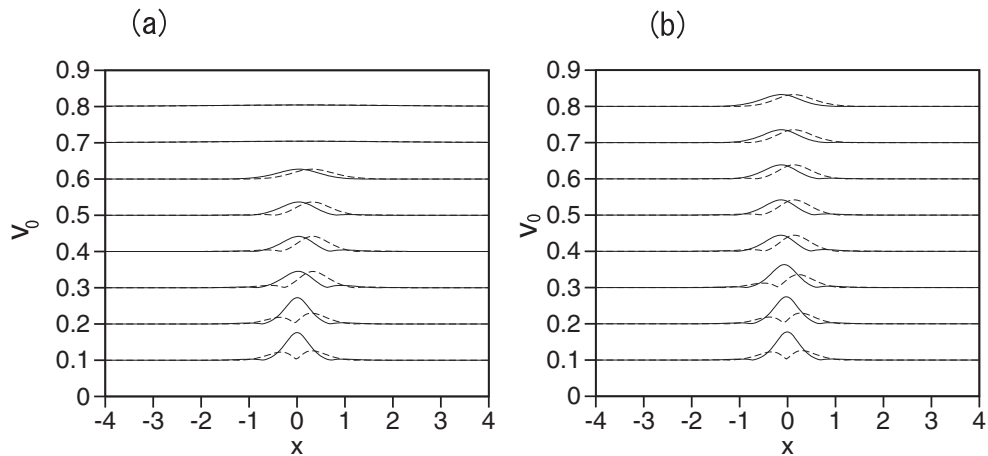}
\end{center}
\caption{(a) Solid and dashed lines show cross-section plots of $|\protect%
\phi _{+}(x)|$ and $|\protect\phi _{-}(x)|$ at $y=0$ for 2D solitons moving
in the free space ($U_{0}=0$), under the action of SOC with strength $%
\protect\lambda =3$ in Eqs. (\protect\ref{tilde+}) and (\protect\ref{tilde-}%
). Values of the velocity, $v_{0}$, are marked in the panel. (b) The same
for the solitons dragged by the square-shaped potential (\protect\ref{2D})
with amplitude $U_{0}=0.5$ and velocity $v_{0}$. The results are obtained
for the solitons with fixed norm $N=5$, setting $\protect\gamma =0$ (the
nonlinear cross-interaction is absent).}
\label{fig1}
\end{figure}

\begin{figure}[h]
\begin{center}
\includegraphics[height=4.5cm]{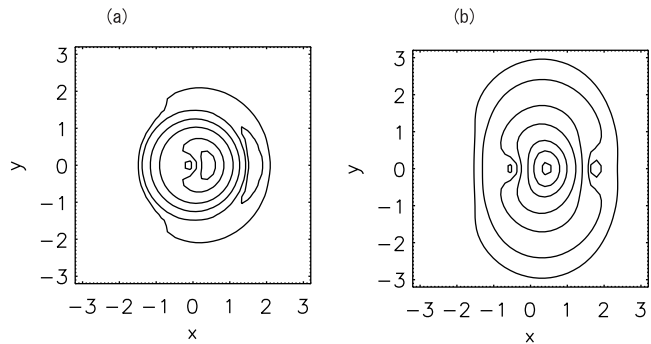}
\end{center}
\caption{Contour plot of $|\protect\phi _{-}(x,y)|$ at $v_{0}=0.2$ (a) and $%
0.4$ (b) for the solitons at $U_{0}=0$. The contour lines are plotted for $|%
\protect\phi _{-}|=0.05,0.1,0.25,0.5,1,1.5,2$.}
\label{fig1d}
\end{figure}

\begin{figure}[h]
\begin{center}
\includegraphics[height=5.5cm]{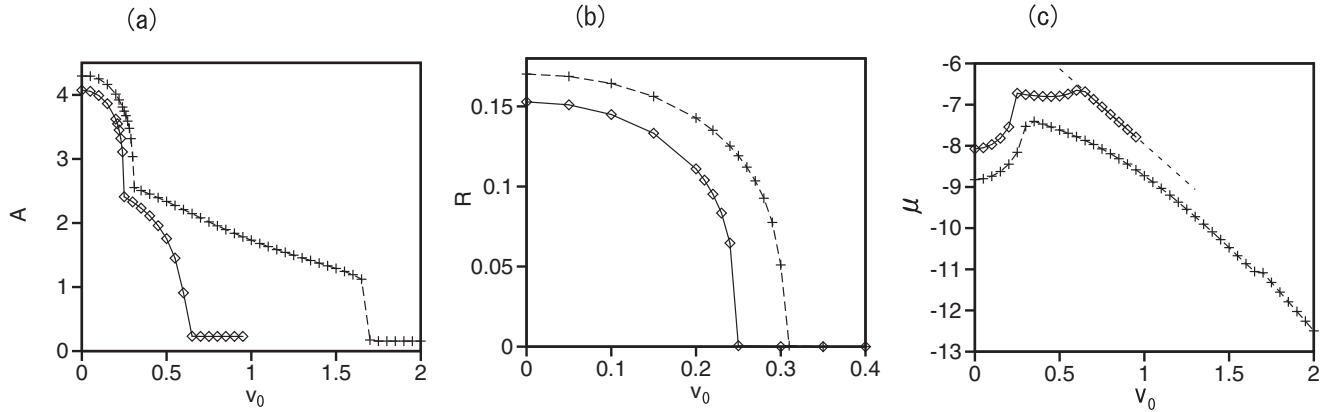}
\end{center}
\caption{(a) Peak values $A$ of squared component $|\protect\phi _{+}|^{2}$
of the moving solitons, as functions of the velocity, $v$, in the free space
($U_{0}=0$, shown by rhombuses), and under the action of the 2D lattice
potential (\protect\ref{tilde}) with $U_{0}=0.5$, shown by crosses. (b) The
asymmetry parameter (\protect\ref{R}) vs. $v$ for the same soliton families.
(c) The chemical potential vs. $v_{0}$ for same families. The dashed line,
plotted as per Eq. (\protect\ref{CW}) with $k_{y}=2\protect\pi \cdot 7/L$,
represents the fully delocalized CW state. The solutions are obtained with $%
\protect\lambda =3$ and $\protect\gamma =0$ in Eqs. (\protect\ref{phi+}) and
(\protect\ref{phi-}). The fixed norm of the solitons is $N=5$.}
\label{fig2}
\end{figure}

\begin{figure}[h]
\begin{center}
\includegraphics[height=5.cm]{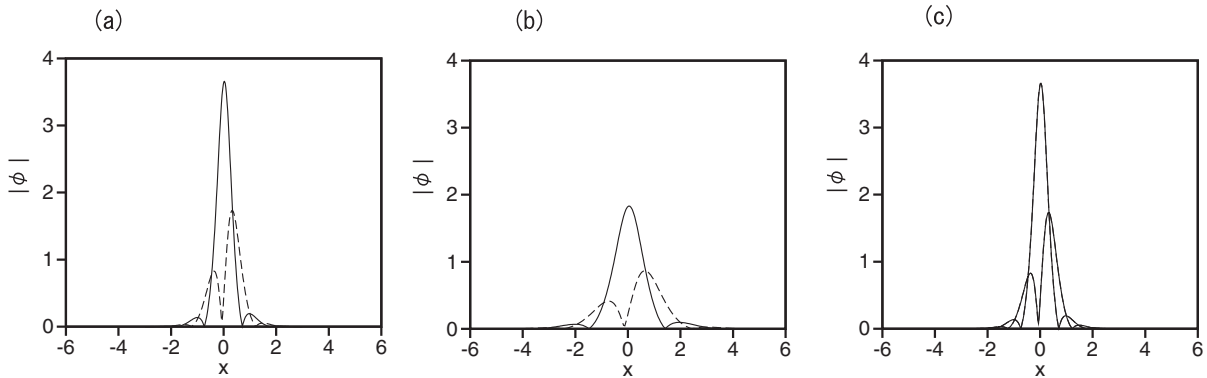}
\end{center}
\caption{(a) Profiles of cross sections $|\protect\phi _{+}(x)|$ and $%
\left\vert \protect\phi _{-}(x)\right\vert $ (solid and dashed lines,
respectively) of the stationary solutions of Eqs. (\protect\ref{phi+2}) and (%
\protect\ref{phi-2}) with $\protect\lambda =3$, $v_{0}=0.2$, and $U_{\mathrm{%
2D}}=0$. (b) The same at $\protect\lambda =1.5$ and $v_{0}=0.1$. (c)
Juxtaposition of $|\protect\phi _{\pm }(x)|$ from (a) (solid lines) with
rescaled profiles $2|\protect\phi _{\pm }^{(s)}(x^{\prime })|$ from (b)
(dashed lines), plotted in rescaled coordinates $x^{\prime }=0.5x,y^{\prime
}=0.5y$. The superimposed profiles completely overlap.}
\label{fig7}
\end{figure}

\begin{figure}[h]
\begin{center}
\includegraphics[height=5.cm]{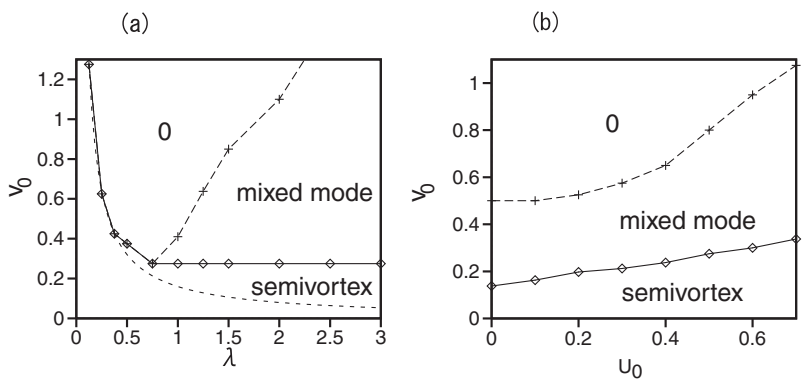}
\end{center}
\caption{(a) Existence regions for SVs and MMs in the plane of the SOC
strength, $\protect\lambda $, and velocity $v_{0}$, in the presence of the
square-shaped potential (\protect\ref{tilde}) with amplitude $U_{0}=0.5$.
Symbol $0$ designates the delocalization area, where solitons do not exist.
The dashed line in (a) is hyperbola $v_{0}=0.16/\protect\lambda $, which
verifies analytical prediction (\protect\ref{approx}), with the fitting
coefficient $0.16$. Panel (b) shows the existence regions in the plane of
the potential's strength, $U_{0}$, and velocity $v_{0}$, for a fixed SOC
strength, $\protect\lambda =1.5$. The results are obtained for $\protect%
\gamma =0$ (no nonlinear interaction between the two components) and the
fixed norm of the solitons, $N=5$.}
\label{fig3}
\end{figure}

\begin{figure}[h]
\begin{center}
\includegraphics[height=4.1cm]{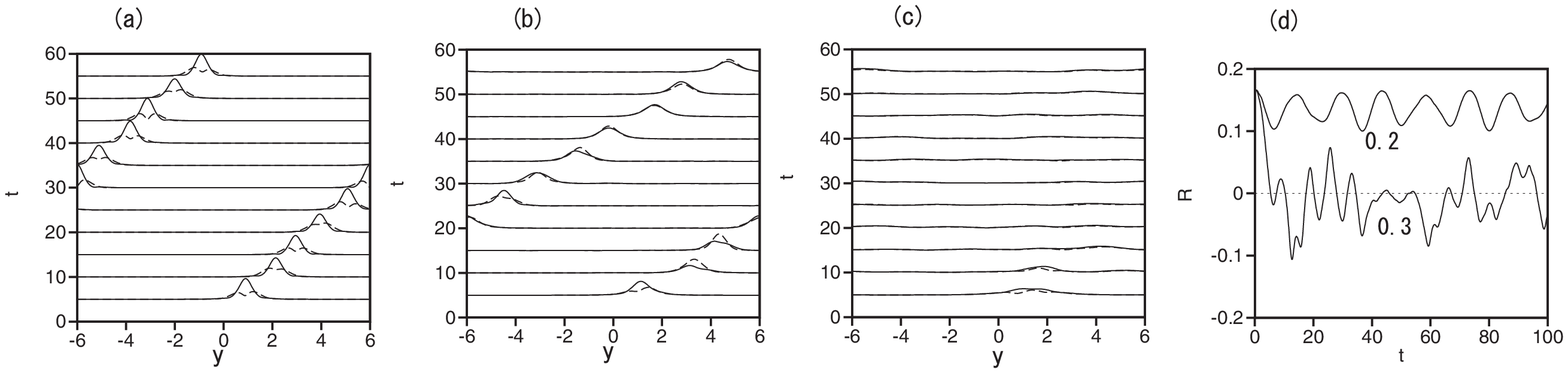}
\end{center}
\caption{(a) Solid and dashed lines display the evolution of $|\protect\phi %
_{+}(y,t)|$ and $|\protect\phi _{-}(y,t)|$ in cross section $x=0$, as
produced by simulations of the propagation governed by Eqs. (\protect\ref%
{phi+}) and (\protect\ref{phi-}) with $\protect\lambda =3$, $\protect\gamma %
=0$, and $U_{0}=0.5$. The input is taken as the stationary soliton with norm
$N=5$ and velocity $\left( v_{0}\right) _{\mathrm{init}}=0.1$, while the
velocity in Eqs. (\protect\ref{phi+}) and (\protect\ref{phi-}) is $\left(
v_{0}\right) _{\mathrm{fin}}=0.2$ in (a), $0.3$ in (b), and $0.5$ in (c).
(d) The evolution of asymmetry parameter (\protect\ref{R}) corresponding to $%
\left( v_{0}\right) _{\mathrm{fin}}=0.2$ and $0.3$, i.e., to panels (a) and
(b), respectively.}
\label{fig4}
\end{figure}

Results for the system with $\gamma =0$ in Eqs. (\ref{phi+}) and (\ref{phi-}%
), when, as mentioned above, only SVs are relevant solutions at $v_{0}=0$,
are presented in Figs. \ref{fig1} -- \ref{fig4}. These results are produced
for the full 2D lattice, defined as per Eq. (\ref{2D}), with amplitude $%
U_{0} $ (some figures display the results for $U_{0}=0$).

First, a set of cross sections of the moving solitons, corresponding to
gradually increasing velocities, are displayed in Fig. \ref{fig1}. Panel (a)
represents, for the sake of comparison, the results for the free space ($%
U_{0}=0$), which corresponds to Ref. \cite{Ben Li}. This set of profiles
demonstrates that, with the increase of $v_{0}$, the SV, which exists at $%
v_{0}=0$, is gradually transformed into a soliton with a quasi-MM structure.
Indeed, linear-mixing terms $\sim \lambda v_{0}$ in the system written in
the form of Eqs. (\ref{tilde+}) and (\ref{tilde-}) make the existence of
pure SVs, whose components carry different vorticities, $0$ and $1$,
impossible. The growth of the mixing terms, with the increase of $v_{0}$,
tends to make the two components mutually mirror-symmetric. The transition
to this shape, which is the signature of the MM structure, occurs in Fig. %
\ref{fig1}(a) at
\begin{equation}
v_{0}=v_{\mathrm{SV\rightarrow MM}}\left( U_{0}=0\right) \approx 0.25.
\label{SVtoMM}
\end{equation}%
Simultaneously, the soliton expands in the $x$ direction, and eventually
disappears, through complete delocalization, at
\begin{equation}
v_{\max }\left( U_{0}=0\right) \approx 0.65.  \label{vmax}
\end{equation}%

Figure \ref{fig1}(b) displays similar results, produced by the
imaginary-time-simulation method in the moving reference frame in the
presence of the square-shaped OL potential (\ref{tilde}) with a relatively
small amplitude, $U_{0}=0.5$. In this case, the SV$\rightarrow $MM and
delocalization transitions occur at, respectively,
\begin{equation}
v_{\mathrm{SV\rightarrow MM}}\left( U_{0}=0.5\right) \approx 0.32,
\label{SVtoMM V0=0.5}
\end{equation}%
\begin{equation}
v_{\max }\left( U_{0}=0.5\right) \approx 1.75.  \label{vmax U0=0.5}
\end{equation}%
cf. Eqs. (\ref{SVtoMM}) and (\ref{vmax}). It is seen that the effect of the
OL potential is small in terms of the former transition, and quite strong
for the expansion of the existence region of the moving solitons. In
addition, Fig. \ref{fig1d} shows the full 2D shape of the solitons by means
of contour plots of $|\phi _{-}(x,y)|$ at $v_{0}=0.2$ (a) and $0.4$ (b) for $%
U_{0}=0$. In particular, the plots clearly show the presence of a
vortex-antivortex pair at $v_{0}=0.4$, which is a characteristic feature of
patterns of the MM type, and is not possible in SVs.

The results for $U_{0}=0$ and $0.5$ are further detailed in Fig. \ref{fig2}.
Panel (a) shows the peak value $A$ of squared component $|\phi _{+}|^{2}$ as
a function of the velocity. In the delocalized state, the peak amplitude
does not fall to zero, because of the finite system's size. Further, the SV$%
\rightarrow $MM transition is quantified in panel (b) by plots of the
parameter characterizing the asymmetry between the two components,%
\begin{equation}
R\equiv N^{-1}\int \int |\phi _{+}\left( x,y\right) |^{2}dxdy-1/2,  \label{R}
\end{equation}%
vs. the velocity for $U_{0}=0$ and $0.5$ (here $N$ is the total norm defined
as per Eq. (\ref{norm})). For MM solitons, whose components are mirror
images of each other, $R=0$, while one has $R>0$ for SVs. Figure \ref{fig2}%
(b) clearly demonstrates the SV$\rightarrow $MM transition at points (\ref%
{SVtoMM}) and \ref{SVtoMM V0=0.5}) for $U_{0}=0$ and $0.5$, respectively.
Further, Fig. \ref{fig2}(c) shows the relationship between chemical
potential $\mu $ of the solitons (see Eq. (\ref{umu})) and velocity $v_{0}$
for the same soliton families. The dashed line is the chemical potential of
the delocalized CW state at $U_{0}=0$, as given by Eq. (\ref{CW}), in which
constant $A^{2}$ is expressed in terms of $N$, and $k_{y}=2\pi \cdot 7/L$ is
chosen. The chain of rhombuses coinciding with the CW line at $v_{0}>v_{\max
}\left( U_{0}=0\right) \approx 0.65$ (see Eq. (\ref{vmax})) represents fully
delocalized states.

To verify the above-mentioned scaling which links different soliton
solutions of Eqs. (\ref{phi+2}) and (\ref{phi-2}) with $U_{0}=0$, we note
that, if a spinor wave function $\phi _{\pm }(x,\tilde{y},t)$ is a solution
for parameters $\left\{ \lambda ,v_{0},\gamma \right\} $, then a solution
for the set of $\left\{ s\lambda ,sv_{0},\gamma \right\} $ is given by
\begin{equation}
\phi _{\pm }^{(s)}=s\phi _{\pm }(sx,s\tilde{y},s^{2}t;s\lambda ,sv_{0}),
\label{r}
\end{equation}%
where $s$ is an arbitrary scaling factor. To check this property, Fig. \ref%
{fig7} shows $|\phi _{+}|$ and $\left\vert \phi _{-}\right\vert $ (solid and
dashed lines, respectively) for the numerically found soliton solutions with
(a) $\left\{ \lambda =3,v_{0}=0.2\right\} $ and (b) $\left\{ \lambda
=1.5,v_{0}=0.1\right\} $, which corresponds to $s=0.5$ in Eq. (\ref{r}). To
check relation (\ref{r}) in detail, Fig. \ref{fig7}(c) compares $|\phi _{\pm
}|$ (with both components drawn by solid lines) for the former solution and
the rescaled version of the latter one, $2\left\vert \phi _{2\pm
}\right\vert $ (dashed lines), plotted in rescaled coordinates, $x^{\prime
}=0.5x,y^{\prime }=0.5y$. The overlap of the profiles confirms scaling
relation (\ref{r}) and, consequently, the linear relation,
\begin{equation}
\left( v_{0}\right) _{\max }(s\lambda )=s\left( v_{0}\right) _{\max
}(\lambda )  \label{lin}
\end{equation}%
for $U_{0}=0$.

As seen below in Figs. \ref{fig3}(b) and \ref{fig6}(b), the presence of the
moving lattice breaks the exact linearity of Eq. (\ref{lin}), but keeps it
as an approximate dependence between $\left( v_{0}\right) _{\max }$ and $%
\lambda $. The slope of the approximately linear dependence is strongly
affected by the lattice -- actually, helping to expand the existence domain
of the moving solitons.

The findings produced by the numerical solution for the solitons dragged by
the 2D lattice are summarized in Fig. \ref{fig3} by diagrams which display
existence regions of the 2D solitons of the SV and MM types in the parameter
plane of $\left( \lambda ,v_{0}\right) $ for $U_{0}=0.5$ (a), and in the
plane of $\left( U_{0},v_{0}\right) $ for a fixed value of the SOC strength,
$\lambda =1.5$ (b). In these plots, symbol $0$ implies the delocalization
(nonexistence of solitons). The MM area appears in Fig. \ref{fig3}(a) at $%
\lambda >0.75$. Again, these plots demonstrate that the effect of the OL
potential is weak for the SV$\rightarrow $MM transition, and strong for the
expansion of the solitons' existence range.

The shape of the left boundary between the SV and $0$ areas in Fig. \ref%
{fig3}(a) is qualitatively explained by Eq. (\ref{approx}), as shown by the
dashed hyperbola, drawn with a fitting coefficient $0.16$. As said above,
this approximation is relevant only for small values of the SOC strength, $%
\lambda $, therefore it does not apply to other boundaries in Figs. \ref%
{fig3}(a) and (b). On the other hand, the roughly linear shape of the
MM-delocalization boundaries is explained, as mentioned above, by the
scaling relation (\ref{linear}).

To verify robustness of solutions for the solitons pulled by the moving OL,
we have also performed direct real-time simulations of equations (\ref{phi+}%
) and (\ref{phi-}) written in the laboratory reference frame. As initial
conditions, we used stationary solutions which were obtained, as above, by
means of the imaginary-time integration in the coordinates moving at a
certain velocity, $\left( v_{0}\right) _{\mathrm{init}}$, while Eqs. (\ref%
{phi+}) and (\ref{phi-}) were simulated in real time with the 2D potential
moving at a higher (final) velocity, $\left( v_{0}\right) _{\mathrm{fin}%
}>\left( v_{0}\right) _{\mathrm{init}}$.

As a result, one might expect, in principle, to observe nonstationary
solitons traveling at some mean speed $\left\langle v\right\rangle <\left(
v_{0}\right) _{\mathrm{fin}}$, so that they lag behind the dragging OL.
However, our simulations have not produced such solutions. Instead, in all
cases when $\left( v_{0}\right) _{\mathrm{fin}}$ belongs to the stability
areas shown in Fig. \ref{fig3}, the initial solitons either pick up the
speed $\left( v_{0}\right) _{\mathrm{fin}}$, moving with some internal
vibrations, or suffer destruction. Characteristic examples are displayed in
Fig. \ref{fig4}. As expected according to Fig. \ref{fig3}(b), the initial
soliton, corresponding to $\left( v_{0}\right) _{\mathrm{init}}=0.1$, and
the established one, with $\left( v_{0}\right) _{\mathrm{fin}}=0.2$ in panel
(a), belong to the SV type, while the soliton eventually traveling at $%
\left( v_{0}\right) _{\mathrm{fin}}=0.3$ in (b) is of the MM type. Finally,
setting $\left( v_{0}\right) _{\mathrm{fin}}=0.5$ in (c) leads to
destruction of the soliton (delocalization), due the large mismatch between $%
\left( v_{0}\right) _{\mathrm{fin}}$ and $\left( v_{0}\right) _{\mathrm{init}%
}$. Residual intrinsic vibrations of the established solitons are
illustrated by oscillations of asymmetry factor $R(t)$ (defined above in Eq.
(\ref{R})), which are displayed in Fig. \ref{fig4}(d). The oscillations are
nearly regular, keeping $R>0$ (i.e., the soliton keeps the asymmetry between
its components, which is a signature of SVs) for $\left( v_{0}\right) _{%
\mathrm{fin}}=0.2$, or irregular, oscillating around $R=0$, for the MM
observed at $\left( v_{0}\right) _{\mathrm{fin}}=0.3$. The regular
oscillations in the case of the semivortex are, essentially, performed by
the "lighter" vortex component of the SV moving around the \textquotedblleft
heavier" zero-vorticity one. The compound soliton of the mixed-mode (MM)
type actually has a larger number of effective degrees of freedom, as its
both components have equal "masses" (norms), and each component features
sub-units with zero and nonzero vorticities. Therefore, the structure of the
MM soliton opens a way to observe more complex internal dynamics.

\subsection{Dragging and steering solitons by quasi-1D potentials}

\begin{figure}[h]
\begin{center}
\includegraphics[height=6.cm]{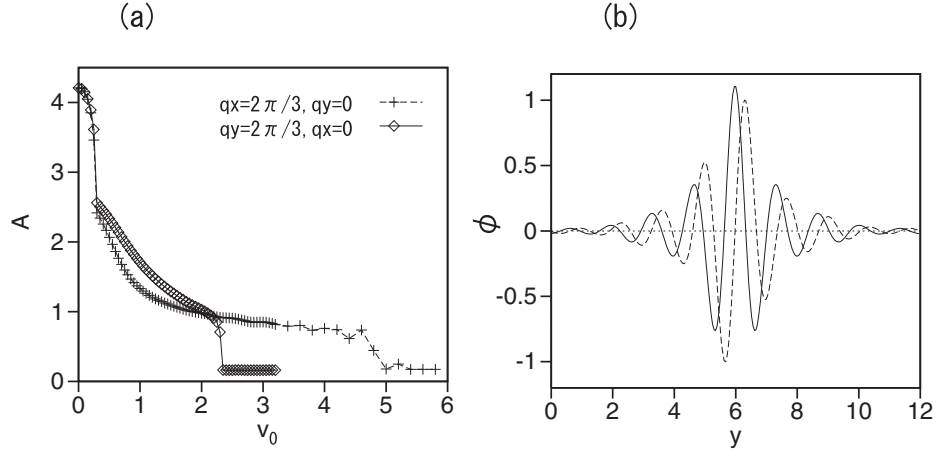}
\end{center}
\caption{(a) The same as in Fig. \protect\ref{fig2}(a) with $U_{0}=0.5$, but
under the action of the quasi-1D potentials (\protect\ref{tilde1D}. The data
labeled \textquotedblleft $qx=2\protect\pi /3,qy=0$" and \textquotedblleft $%
qy=2\protect\pi /3,qx=0$" correspond, respectively, to potentials $U_{%
\mathrm{1D}}(x)$ and $U_{\mathrm{1D}}(\tilde{y})$. (b) Snapshot profiles of
the wave-function components Re$\left( \protect\phi _{+}\right) $ and Im$%
\left( \protect\phi _{+}\right) $, in cross section $x=0$, for the soliton
with $v=1.5$, pulled by the moving potential $U_{\mathrm{1D}}(\tilde{y})$.}
\label{fig5}
\end{figure}

Proceeding to results obtained from Eqs. (\ref{tilde+}) and (\ref{tilde-})
with $U_{\mathrm{2D}}$ replaced by quasi-1D potentials (\ref{tilde1D}), Fig. %
\ref{fig5}(a) shows the corresponding characteristics of the soliton
families defined as in Fig. \ref{fig2}, i.e., the peak value $A$ of $|\phi
_{+}|^{2}$ as a function of $v_{0}$. First, we notice that, if the 2D
potential is replaced by $U_{\mathrm{1D}}(\tilde{y})$, with the same
amplitude and wavenumber as above, $U_{0}=0.5$ and $q=2\pi /3$, the largest
velocity, up to which the quasi-1D potential can drag the soliton, increases
from the value given by Eq. (\ref{vmax U0=0.5}) to $v_{\max }\approx 2.25$.

A characteristic example of the longitudinal structure of the stable soliton
dragged by potential $U_{\mathrm{1D}}(\tilde{y})$ is displayed in Fig. \ref%
{fig5}(b) (in the transverse $x$ direction, the soliton features a smooth
localized shape). A dominant wavenumber related to this multi-peak structure
is $k_{y}\simeq 4.6$. It is relevant to note that this value is very
different from $q/2=\pi /3\approx 1.05$ (see Eq. (\ref{q})), which is one
that may be, in principle, singled out by the condition of the parametric
resonance induced by terms $U_{0}\cos \left( q\tilde{y}\right) \tilde{\phi}%
_{\pm }$ in Eqs. (\ref{tilde+}) and (\ref{tilde-}), cf. Eqs. (\ref{ab}) and (%
\ref{res}). Thus, the dynamical mechanisms considered in this work are not
affected by the possibility of the resonance.

On the other hand, if the quasi-1D potential is taken as $U_{\mathrm{1D}}(x)$
in Eq. (\ref{tilde1D}), once again with the same amplitude and wavenumber as
above, the highest velocity admitting stable motion of the solitons (which
are actually not dragged, but steered along the guiding channel, in such a
case) is much larger. Indeed, Fig. \ref{fig4}(a) produces $v_{\max }\approx 5
$ for the solitons traveling under the action of potential $U_{\mathrm{1D}%
}(x)$. This value is, roughly, \emph{three times} larger than its
counterpart (\ref{vmax U0=0.5}) obtained above in the case of the 2D
potential. The steep increase of $v_{\max }$ in the latter case is explained
by the fact that the potential $U_{\mathrm{1D}}(x)$ tends to compress the 2D
system of Eqs. (\ref{tilde+}) and (\ref{tilde-}) into its 1D version. In
turn, the 1D system may be reduced, as mentioned above, to the single GPE ( %
\ref{single}), which has no limitation for the existence of solitons at any
velocity.

The increase of $v_{\max }$ following the replacement of the 2D lattice by
the quasi-1D ones has also been checked for the lattice wavenumbers
different from value (\ref{q}) which was fixed above. It was thus found that
\begin{eqnarray}
q &=&5\pi /6:v_{\max }^{\mathrm{(2D)}}=1.225,v_{\max }^{\mathrm{(1D,}y%
\mathrm{)}}=1.525,v_{\max }^{\mathrm{(1D,}x\mathrm{)}}=3.175,  \label{5/6} \\
q &=&2\pi /3:v_{\max }^{\mathrm{(2D)}}=1.75,v_{\max }^{\mathrm{(1D,}y\mathrm{%
)}}=2.25,v_{\max }^{\mathrm{(1D,}x\mathrm{)}}\approx 5,  \label{2/3} \\
q &=&\pi /2:v_{\max }^{\mathrm{(2D)}}\approx 3.2,v_{\max }^{\mathrm{(1D,}y%
\mathrm{)}}\approx 5.4,v_{\max }^{\mathrm{(1D,}x\mathrm{)}}\approx 13.35,
\label{1/2}
\end{eqnarray}%
respectively, for $U_{\mathrm{2D}}(x,y)$, $U_{\mathrm{1D}}(y)$, and $U_{%
\mathrm{1D}}(x)$. Here, for the sake of comparison, Eq. (\ref{2/3})
reproduces the above-mentioned results for $q=2\pi /3$. It is seen that, in
all the cases, $v_{\max }$ decreases with the increase of $q$. This trend is
naturally explained by the fact that the convolution of the soliton's wave
function with the rapidly oscillating OL potential produces a weaker effect.

\subsection{Dragging MM solitons in the presence of the nonlinear
cross-interaction ($\protect\gamma >1$)}

As said above, the quiescent ($v_{0}=0$) ground-state solutions of Eqs. (\ref%
{phi+}) and (\ref{phi-}) with $\gamma >1$ in the free space ($V_{0}=0$) are
MM solitons, while the SVs are unstable in this case. Then, the 2D potential
(\ref{tilde}) can drag the MMs up to the respective limit velocity, $v_{\max
}$, above which the solitons suffer delocalization. First, the boundary
between the moving MMs and delocalized states in the free space ($U_{0}=0$)
is plotted in Fig. \ref{fig5}(a) for the system with $\gamma =2$. The linear
shape of the boundary is rigorously explained by Eq. (\ref{linear}) which
follows from the scaling properties of Eqs. (\ref{phi+}) and (\ref{phi-})
with $U_{0}=0$ and a fixed norm. Next, the same boundary, but in the
presence of the dragging potential, is plotted in Fig. \ref{fig5}(b). It is
seen that even relatively weak potentials, with $U_{0}=0.2$ and $0.4$, help
to strongly expand the stability area for the moving MM solitons.
\begin{figure}[h]
\begin{center}
\includegraphics[height=6.cm]{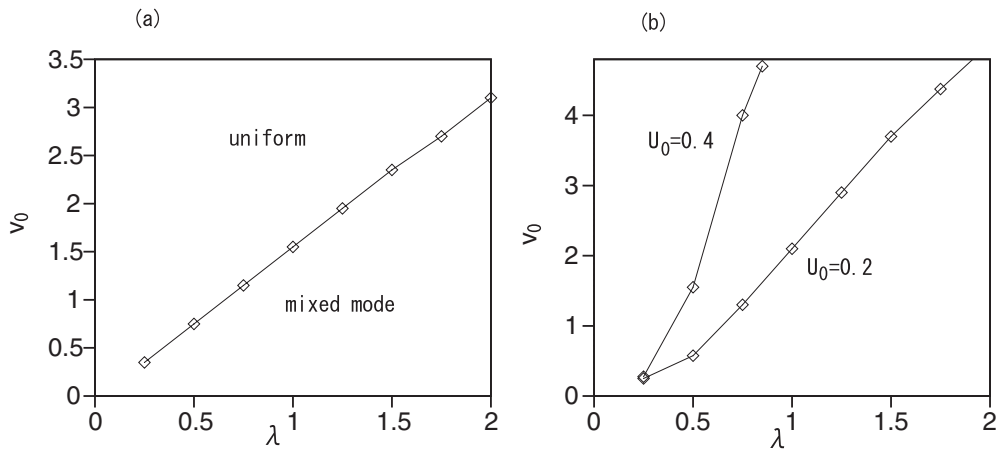}
\end{center}
\caption{(a) The boundary between the MMs moving in the free space ($U_{0}=0$%
) and the delocalized (uniform) state, in the plane of $\left( \protect%
\lambda ,v\right) $. (b) The same, in the presence of the 2D potential (%
\protect\ref{tilde}) with $U_{0}=0.2$ and $0.4$. The data are presented for
the system with $\protect\gamma =2$ and a fixed norm of the MM solitons, $N=3
$.}
\label{fig6}
\end{figure}

\section{Conclusion}

In this work, we have demonstrated that mobility limits for two-component 2D
matter-wave solitons, stabilized by the SOC effect, may be strongly expanded
by means of relatively weak spatially periodic potentials moving at a
desirable speed. Boundaries of the stable motion are identified by means of
numerical methods, and the shape of some boundaries is explained
analytically. If the stable quiescent solitons are SVs (semivortices), the
motion converts them into MMs (mixed modes), which suffer delocalization at
much higher velocities. A remarkable finding is that quasi-1D potentials,
especially the one with the wave vector directed perpendicular to the
velocity, provide essentially stronger stabilization than the full 2D
lattice.

The analysis reported here may be extended to develop schemes for the
transfer of a soliton by a moving potential from an initial position to a
predetermined final one, cf. Refs. \cite{transport1,transport2,transport-SOC}%
. Further, it may be interesting to develop the analysis for two- and
multi-soliton complexes trapped in the lattice potential. A challenging
option, suggested by Ref. \cite{Pu}, is to consider a possibility to
stabilize 3D moving solitons, which are made metastable by SOC in the
quiescent state.

\section*{Acknowledgments}

This was supported, in part, by the Israel Science Foundation though grant
No. 1286/17, and by the Japan Society for the Promotion of Science KAKENHI
(Grant No. 18K03462)

\section*{ORCID\ IDs}

B. A. Malomed 0000-0001-5323-1847; H. Sakaguchi 0000-0003-3591-0496

\end{document}